\documentclass[twocolumn,pre]{revtex4}   

\usepackage{dcolumn}
\usepackage{amsmath}

\usepackage{graphicx}
\usepackage{subfigure}

\newcommand{\e}{\mathrm{e}}
\newcommand{\half}{\tfrac12}
\newcommand{\set}[1]{\lbrace#1\rbrace}
\newcommand{\etal}{{\it{}et~al.}}
\newcommand{\defn}{\textit}
\newcommand{\mat}{\mathbf}
\renewcommand{\vec}{\mathbf}
\newcommand{\Ord}{\mathrm{O}}

\begin{document}

\title{Estimating network structure from unreliable measurements}
\author{M. E. J. Newman}

\affiliation{Department of Physics and Center for the Study of Complex Systems,\\University of Michigan, Ann Arbor, MI 48109, USA}

\begin{abstract}
Most empirical studies of networks assume that the network data we are given represent a complete and accurate picture of the nodes and edges in the system of interest, but in real-world situations this is rarely the case.  More often the data only specify the network structure imperfectly---like data in essentially every other area of empirical science, network data are prone to measurement error and noise.  At the same time, the data may be richer than simple network measurements, incorporating multiple measurements, weights, lengths or strengths of edges, node or edge labels, or annotations of various kinds.  Here we develop a general method for making estimates of network structure and properties using any form of network data, simple or complex, when the data are unreliable, and give example applications to a selection of social and biological networks.
\end{abstract}
\maketitle

\section{Introduction}
Networks are widely used as a convenient quantitative representation of patterns of connection between the nodes, units, agents, or elements in complex systems, particularly technological, biological, and social systems.  There has been an explosion of empirical work in the last two decades aimed at measuring and describing the structure of networks such as the Internet, the World Wide Web, road and airline networks, friendship networks, biochemical networks, ecological networks, and others~\cite{Newman18c}.

A fundamental issue with empirical studies of networks, however, is that the data we have are often unreliable.  Most measurement techniques for network structure suffer from measurement error of some kind.  In biological networks such as metabolic or protein interaction networks, for example, traditional laboratory experimental error is a primary source of inaccuracy.  There exist experimental methods for directly measuring interactions between proteins, for instance, such as affinity purification or yeast two-hybrid screens, but even under the best controlled conditions the exact same measurement repeated twice may yield different results~\cite{Ito01,Krogan06,Rolland14}.  Or consider the Internet, whose network structure is usually determined by examining either router tables or collections of trace\-route paths.  Both router tables and traceroute paths, however, give only subsets of the edges in the network.  Commonly one combines many tables or paths to provide better coverage, but even so it is well established that the resulting network structure contains significant errors~\cite{LBCX03,ACKM05,CM05}.  Social networks, such as friendship networks, provide another example.  Such networks are typically measured using surveys, and the resulting data can contain errors of many kinds, including subjectivity on the part of respondents in surveys, missing data, and recording and coding errors~\cite{KB76,Marsden90,Butts03,HG10,LKR12}.

Not all is bad news, however.  Network data may be error-prone but they can also be very rich.  Many studies produce not just simple measurements of network structure but multifaceted data sets that reflect structure from many different angles.  A friendship network might be measured repeatedly, for instance, or measured in multiple ways using reported interactions, observed interactions, online data, or archival records.  An ecological network such as a food web might combine field data of many different types as well as input from curated library studies.  A data set for the World Wide Web will typically include not only the pattern of links between web pages but rich data on page content including word frequencies, headings, word positions, anchor text, and metadata.

In this paper we consider the problem of \defn{network reconstruction}, deriving and demonstrating a broad class of methods that can be used to infer the true structure and properties of a network from potentially error-prone data in any format.

There has been a significant amount of previous work on network error and reconstruction, including discussion of the types of errors that can occur, for instance in social networks~\cite{KB76,Marsden90,Butts03,HG10,LKR12}, biological networks~\cite{Krogan06,WPVS09}, and technological networks~\cite{LBCX03,ACKM05,CM05}, and studies of simulated errors that aim to determine what effect errors will have on estimates of network properties~\cite{FC02,BCK06,Kossinets06,KLN08,WSML12,ET15}.  Methods for estimating true network structure from error-prone measurements have been developed in several fields, including sociology, statistics, physics, and computer science.  Perhaps most closely related to the work reported here is that of Butts~\cite{Butts03}, who developed Bayesian methods for estimating how reliable social network survey respondents are in the answers they give.  Our technical approach is different from that of Butts, but some of the example applications we consider address similar questions.  The methods we develop generate a posterior distribution over possible network structures, and a number of other previous methods have been proposed for doing this, by ourselves and others, albeit using different approaches~\cite{CMN08,GS09}.  Looking further afield, there is also a considerable body of work on estimating network structure from measurements of the evolution of networked dynamical systems such as coupled oscillators~\cite{WLG16} or spreading processes~\cite{LMOZ14,LL17}.  There has also been work on error correction strategies for networks, which can be viewed as a form of network reconstruction.  Link prediction in particular---the task of identifying missing edges in networks---has received considerable attention~\cite{LK07,CMN08,GS09,Huisman09,KL11}.  In the study of citation and collaboration networks a number of methods have been developed for name disambiguation, which can be thought of as a form of error correction for missing or extraneous nodes~\cite{ST09,DGA11,FGL12,TFWZ12,MBKN13}.  And a combination of methods of these kinds can be used to create hybrid ``coupled classifier'' algorithms for processing raw network data into more accurate and nuanced estimates of network quantities~\cite{NKG11,HSWD15,CNSS17,Angulo17}, often with a focus on a specific domain of study.  Of particular interest is the large body of work on methods for analyzing and processing high-throughput laboratory data on biological and biochemical networks~\cite{Forster03,LLZ05,SS05,Margolin06,LH08b,Allen12}.

In a previous paper~\cite{Newman18a} we outlined a method for making optimal estimates of network structure using expectation--maximization (EM) algorithms, and in other work we and others have looked at methods for estimating networks in the presence of community structure~\cite{MBN16,LLL18}.  Here we build on this previous work and lay out a general formalism for inferring network structure from rich but noisy data sets.  We focus specifically on the problem of inferring the positions of the edges in a network of given nodes.  There are interesting questions to be asked about how one identifies the nodes in the first place, but these we will not tackle here.

\section{Approach}
\label{sec:em}
The approach we present, which builds on our previous work in~\cite{MBN16,Newman18a}, is based on model fitting and has two main components: a~\defn{network model} that represents how the network structure is generated (or, more properly, our belief about how it is generated), and a \defn{data model} that represents how that structure maps onto the observed data.  Given a set of observations, the method allows us to infer the parameters of both models as well as the entire posterior distribution over possible network structures.  Features of interest in the network can also be estimated, in one of two different ways: one can inspect the parameters of the network model (as is done in community detection, for example) or one can calculate expected values of network metrics over the posterior distribution (as one might for things like degree distributions, path lengths, or correlations).

The requirement that we define network and data models is one aspect that distinguishes our method from other network reconstruction approaches.  One might consider this requirement to be a disadvantage of the method, since it obliges us to make assumptions about our networks, but we would argue that this is a feature, not a bug.  We argue that other methods are also making assumptions, though one may not notice them because they are often hidden from view.  They may be implicit, for instance, in the decisions a programmer makes in developing code, or they may be the result of subconscious choices not even recognized by the researcher, but they nonetheless still affect the calculations.  We believe it to be a desirable feature of the approach proposed here that it obliges us to acknowledge explicitly what assumptions we are making and formulate them in a precise manner.

Suppose then that we are interested in a particular network of $n$ nodes, whose structure we will represent by an adjacency matrix~$\mat{A}$.  In the simplest case of an unweighted undirected network the adjacency matrix is an $n\times n$ symmetric matrix with elements $A_{ij}=1$ if nodes $i$ and~$j$ are connected by an edge and 0 otherwise.  Our methods can also be applied to directed networks (represented by asymmetric matrices), weighted networks (represented by matrices containing values other than 0 and~1), and other more complicated forms if necessary, but for the moment we will concentrate on the undirected unweighted case.

We assume that the structure~$\mat{A}$ of the network is initially unknown.  This structure is sometimes called the \defn{ground truth}.  Our aim is to estimate the ground truth from the results of measurements of some kind.    The measurements could take many forms: measurements of single edges, pathways, or subgraphs; repeated measurements or measurements made from the point of view of different participants or locations; metadata concerning edges or nodes; nonlocal or global properties of the network as a whole, such as densities, clustering coefficients, or spectral properties, or any of many other measurement types.  Let us denote by~$D$ the complete set of data generated by the measurements performed on the system.  We specifically do not assume that the data are reliable (they may contain errors of various kinds) or that they are complete (some parts of the network may not be measured).  Our goal is to make the best estimate we can of the structure of the network given the available data.

This we do using probabilistic methods.  We first define a network model, which represents our prior knowledge about the structure of the network.  This model takes the form of a probability distribution~$P(\mat{A}|\gamma)$, where~$\gamma$ denotes the parameters (if any) of the distribution.  The parameters are normally unknown; we will show how to estimate their values shortly.  The network model quantifies what we know about the structure of the network before we observe any data.  The model could be a very simple one.  For instance, in the (common) case where we know nothing about the structure of the network ahead of time, the model could be just a uniform (maximum-entropy) distribution, all structures being equally likely.  In fact, we usually know at least a little more than this.  For instance, almost all empirically observed networks are relatively sparse, meaning that only a small fraction of their possible edges are present.  Armed with this additional knowledge, we might choose to employ a network model such as a random graph, that can model networks with low edge density.  We use models of this kind in several calculations in this paper.  More complex choices are also possible and may be useful in some cases.  If we are interested in performing community detection on our network, for example, then we might hypothesize that the network is drawn from a stochastic block model~\cite{HLL83}.  The parameters of the fitted block model can then tell us about the community structure, if any, in the observed network~\cite{BC09,KN11a,MBN16,LLL18}.

Second, we hypothesize a measurement process or data model that describes how our empirical data~$D$ are generated from observations of the network, such that $P(D|\mat{A},\theta)$ is the probability of the data given the true structure of the network~$\mat{A}$ and model parameters~$\theta$.  Combining probabilities and applying Bayes rule, we then have
\begin{equation}
P(\mat{A},\gamma,\theta|D) = {P(D|\mat{A},\theta) P(\mat{A}|\gamma)
  P(\gamma) P(\theta)\over P(D)},
\label{eq:bayes}
\end{equation}
where $P(\gamma)$, $P(\theta)$, and~$P(D)$ are the prior probabilities of the parameters and the data (which we assume to be independent).  Summing over all possible values of the unknown adjacency matrix~$\mat{A}$ (or integrating in the case of continuous-valued matrix elements), we get an expression for the posterior probability of the parameter values~$\gamma,\theta$ given the data:
\begin{equation}
P(\gamma,\theta|D) = \sum_\mat{A} P(\mat{A},\gamma,\theta|D).
\label{eq:posterior}
\end{equation}
Our first goal will be to find the most likely values of the parameters by maximizing this posterior probability with respect to~$\gamma$ and~$\theta$, a so-called \defn{maximum a posteriori} or MAP estimate.

In fact, as is often the case, it is convenient to maximize not the probability itself but its logarithm, which has its maximum in the same place.  We make use of Jensen's inequality, which states that for any set of positive quantities~$x_i$,
\begin{equation}
\log \sum_i x_i \ge \sum_i q_i \log {x_i\over q_i},
\end{equation}
where $q_i$ are an equal number of nonnegative quantities satisfying $\sum_i q_i = 1$.  Applying this inequality to the log of Eq.~\eqref{eq:posterior} we have
\begin{align}
\log P(\gamma,\theta|D) &= \log \sum_\mat{A} P(\mat{A},\gamma,\theta|D)
  \nonumber\\
  &\ge \sum_\mat{A} q(\mat{A}) \log {P(\mat{A},\gamma,\theta|D)\over
                                     q(\mat{A})},
\label{eq:ineq}
\end{align}
where $q(\mat{A})$ is any nonnegative function of~$\mat{A}$ satisfying $\sum_{\mat{A}} q(\mat{A}) = 1$.  It will be convenient to think of $q(\mat{A})$ as a probability distribution over networks~$\mat{A}$.

It is straightforward to see that the exact equality in~\eqref{eq:ineq} is achieved, and hence the right-hand side of the inequality maximized, when
\begin{equation}
q(\mat{A}) = {P(\mat{A},\gamma,\theta|D)\over
              \sum_\mat{A} P(\mat{A},\gamma,\theta|D)}.
\label{eq:estep}
\end{equation}
Since this choice makes the right-hand side equal to $\log P(\gamma,\theta|D)$, a further maximization with respect to $\gamma$ and~$\theta$ will then give us the MAP estimate that we seek.  To put that another way, maximization of the right-hand side of~\eqref{eq:ineq} with respect both to~$q$ and to $\gamma$ and~$\theta$ will give us the optimal values of the parameters.

This leads to a natural iterative algorithm for determining the values of the parameters: we perform the maximization by maximizing first over~$q$ with the parameters held constant, then over the parameters with $q$ held constant, and repeat until we converge to the final answer.  The maximum over~$q$ is given by Eq.~\eqref{eq:estep}.  The maximum over the parameters we find by differentiating.  Taking derivatives of the right-hand side of Eq.~\eqref{eq:ineq} while holding $q(\mat{A})$ constant, we get
\begin{align}
\label{eq:mstepa1}
\sum_{\mat{A}} q(\mat{A}) \nabla_\gamma \log P(\mat{A},\gamma,\theta|D) &= 0, \\
\label{eq:mstepa2}
\sum_{\mat{A}} q(\mat{A}) \nabla_\theta \log P(\mat{A},\gamma,\theta|D) &= 0,
\end{align}
where $\nabla_\gamma,\nabla_\theta$ denote derivatives with respect to the sets $\gamma,\theta$ of parameters of the two models.  Alternatively, making use of Eq.~\eqref{eq:bayes}, we have
\begin{align}
\label{eq:mstep1}
\nabla_\gamma \log P(\gamma) + \sum_{\mat{A}} q(\mat{A}) \nabla_\gamma \log P(\mat{A}|\gamma) &= 0, \\
\label{eq:mstep2}
\nabla_\theta \log P(\theta) + \sum_{\mat{A}} q(\mat{A}) \nabla_\theta \log P(D|\mat{A},\theta) &= 0.
\end{align}
The solution of these equations gives us our values for~$\gamma,\theta$.  Note that Eq.~\eqref{eq:mstep1} depends only on the network model and its solution gives the parameter values for that model.  Similarly, Eq.~\eqref{eq:mstep2} depends only on the data model and gives the parameters for that model.

This is an example of an expectation--maximization or EM algorithm~\cite{DLR77,MK08}, a standard tool for statistical inference in situations where some data are unknown or hidden from us---in this case the network structure~$\mat{A}$.  Implementation of the algorithm involves choosing random initial values for the parameters~$\gamma,\theta$ and then iterating Eq.~\eqref{eq:estep} and Eqs.~\eqref{eq:mstep1} and~\eqref{eq:mstep2} until convergence is reached.  The EM algorithm can be proved to converge to a local maximum of the posterior probability, but not necessarily to the global maximum we would like to find.  In practice, therefore, one often performs repeated runs, starting from different initial values, to test for consistent convergence.

The output of the EM algorithm is a set of values for the parameters~$\gamma,\theta$.  These are ``point estimates,'' representing the single most likely values for the quantities in question.  There exist other (Bayesian) methods that can compute entire posterior distributions over parameters but in most of the applications we consider such an approach is unnecessary.  The quantity of data embodied in the networks we study is typically large enough that the parameter values are quite precisely determined, meaning that the posterior distributions are sharply peaked, and hence the EM algorithm tells us everything we want to know.  Just as in traditional statistical mechanics, the fact that we are studying a large system makes the point estimates highly accurate.  (One exception occurs when we use a model that has an extensive number of parameters.  In this case a Bayesian approach may give additional information that cannot be derived from the EM algorithm, but we will not pursue such approaches in this paper.)

After calculating the parameter values the next step would normally be to use them in Eq.~\eqref{eq:bayes} to find the probability distribution over networks~$\mat{A}$.  It turns out, however, that this is unnecessary, since the network structure can be deduced from results we have already calculated.  Note that Eq.~\eqref{eq:estep} can be written as
\begin{equation}
q(\mat{A}) = {P(\mat{A},\gamma,\theta|D)\over P(\gamma,\theta|D)}
           = P(\mat{A}|D,\gamma,\theta).
\end{equation}
In other words, $q(\mat{A})$~is the probability that the network has structure~$\mat{A}$ given the observed data and our values for the parameters~$\gamma,\theta$.  Thus the EM algorithm already gives us the entire posterior distribution over possible ground-truth network structures.  In many cases this posterior probability distribution is the primary object of interest in the calculation, capturing both the network structure itself and the uncertainty in that structure.

Once we have this distribution, any other network quantity we are interested in, including degrees, correlations, clustering coefficients, and so forth, can be estimated from it.  Specifically, for any quantity $X(\mat{A})$ that is a function of the network structure~$\mat{A}$, the expected value, given the observed data and the parameter estimates, is
\begin{equation}
\mu_X = \sum_\mat{A} X(\mat{A}) P(\mat{A}|D,\gamma,\theta),
\label{eq:mux}
\end{equation}
and the variance about that value is
\begin{equation}
\sigma^2_X = \sum_\mat{A} [ X(\mat{A}) - \mu_X ]^2 P(\mat{A}|D,\gamma,\theta).
\label{eq:sigmax}
\end{equation}
It is not always possible to perform the sums over~$\mat{A}$ in these expressions analytically.  In cases where they cannot be done, numerical approximations using Monte Carlo sampling can give good answers in reasonable time.

The values of the model parameters may also be of interest, both for the network model and for the data model.  In cases where the parameters of the network model correspond to meaningful network quantities, they can give us useful information, as in the case of community detection using the stochastic block model~\cite{MBN16,LLL18}.  More commonly, however, it is the parameters of the data model that are of interest because they quantify the measurement process and hence can give us insight into the reliability of the data and the types of error they may contain.

\section{Network models}
\label{sec:networkmodels}
Applying the methods of the previous section requires us to choose the models we will use: the network model, which describes the prior probability distribution over networks, and the data model, which describes how the data are related to the network structure.  In this and the following section we give some examples of possible choices, starting with network models.

The network models most commonly used for structural inference in networks are random graph models in which the edges are (conditionally) independent random variables.  The best known examples are the (Bernoulli) random graph, the configuration model, the stochastic block model, and their many variants.

\subsection{The random graph}
\label{sec:rg}
The simplest of network models is the standard random graph, in which every pair of distinct nodes~$i,j$ is connected by an edge with equal probability~$\omega$.  For this model the probability~$P(\mat{A}|\gamma)$ becomes
\begin{equation}
P(\mat{A}|\omega) = \prod_{i<j} \omega^{A_{ij}} (1-\omega)^{1-A_{ij}}.
\label{eq:rglike}
\end{equation}
Despite its simplicity, this model works well for many of the calculations we will look at.  In the absence of evidence to the contrary, simply assuming that all edges are equally likely is a sensible approach.  We do need to choose a prior probability~$P(\omega)$ for the parameter~$\omega$.  In the calculations we perform we will assume that all values of this parameter are equally likely, so that $P(\omega)=1$.

\subsection{Edge types}
\label{sec:types}
Various extensions of the simple random graph are possible.  For instance, one could have a model in which instead of just two edge states (present/not present) we have three or more.  In a social network, for instance, one might divide pairs of individuals into those who are not acquainted, somewhat acquainted, or well acquainted.  Such states could be represented by adjacency matrix elements with values 0, 1, and~2, with corresponding probabilities~$\omega_0$, $\omega_1$, and~$\omega_2$.  More generally any number~$k$ of states could be represented by $A_{ij}=0\ldots k-1$ and probabilities $\omega_0\ldots\omega_{k-1}$, subject to the constraint that $\sum_{m=0}^{k-1} \omega_m = 1$.  Then the probability of a particular network is
\begin{equation}
P(\mat{A}|\omega) = \prod_{i<j} \omega_{A_{ij}}
   = \prod_{i<j}\,\prod_{m=0}^{k-1} \omega_m^{\delta_{m,A_{ij}}}.
\label{eq:types}
\end{equation}

A variant of this type of model is one in which the edges are signed, meaning that they can have both positive and negative values.  Such signed networks are sometimes used, for instance, to represent social networks in which interactions can be both positive and negative---friendship and animosity~\cite{WF94}.  In the simplest case the elements of the adjacency matrix take three values $0$, $+1$, and~$-1$, and the probability~$P(\mat{A}|\omega)$ is an obvious variation on Eq.~\eqref{eq:types}.

\subsection{Poisson edge model}
\label{sec:poisson}
In many calculations with network models one assumes not Bernoulli (i.e.,~zero/one) random variables for the edges but Poisson ones.  That is, rather than placing edges with probability~$\omega$ or not with probability~$1-\omega$, one places a Poisson distributed number of edges with mean~$\omega$.  This results in a network that can contain more than one edge between a given pair of nodes---a so-called multiedge---which is in a sense unrealistic since most observed networks do not have multiedges.  However, the probability of having a multi\-edge is of order~$\omega^2$, which is typically negligible in the common case of a sparse network where $\omega$ is small, and hence the Poisson and Bernoulli models generate essentially the same ensemble in the sparse case.  At the same time the Poisson model is often significantly easier to work with and has become favored for many applications.  Commonly, in addition to multiedges, one also allows self-edges, placing a Poisson-distributed number of such edges at each node with mean~$\half\omega$.  By convention a self-edge is represented by an adjacency matrix element~$A_{ii}=2$ (not~1).  The factor of $\half$ in the density of self-edges compensates for the 2 in the definition of~$A_{ii}$, so that the expected value of all adjacency matrix elements is simply~$\omega$.

For this model the equivalent of Eq.~\eqref{eq:rglike} is
\begin{equation}
P(\mat{A}|\omega) = \prod_{i<j} {\omega^{A_{ij}}\over A_{ij}!} \e^{-\omega}
                  \prod_i {\bigl(\half\omega\bigr)^{A_{ii}/2}\over
                  \bigl(\half A_{ii}\bigr)!} \e^{-\omega/2},
\end{equation}
and the log of this probability is
\begin{align}
\log P(\mat{A}|\omega) &= \half \sum_{ij} \bigl( A_{ij} \log \omega - \omega \bigr)
  \nonumber\\
  &{} - \sum_{i<j} \log A_{ij}! - \sum_i \bigl[ \half A_{ii} \log 2
  + \log (\half A_{ii})! \bigr].
\label{eq:logpoisson}
\end{align}
Here we have separated out terms that do not depend on~$\omega$.  When we perform a derivative as in Eq.~\eqref{eq:mstep1}, these terms will vanish, leaving an especially simple result for~$\omega$.

Note also that this model and the previous one both use values of $A_{ij}$ other than zero and one, but the values have different meanings.  In the model of Section~\ref{sec:types} they represent different types of edges; in the model of this section they represent multiedges.

\subsection{Stochastic block model}
\label{sec:sbm}
A more complex model, well studied in the networks literature, is the stochastic block model.  First proposed in the 1980s by Holland~\etal~\cite{HLL83}, the stochastic block model is a model of community structure in networks, although with large numbers of blocks it can also function as a model of very general kinds of network structure~\cite{Borgs08,Lovasz12}.  In essence, the model consists of a set of random graphs stitched together into a larger network.  We take $n$ nodes and divide them into some number~$k$ of groups labeled by integers $1\ldots k$, with $\mu_r$ being the probability that a node is assigned to group~$r$.  Then we place undirected edges between distinct nodes independently such that the probability of an edge between a given pair of nodes depends only on the groups that the nodes belong to.

The model is simplest when written using the Poisson formulation of Section~\ref{sec:poisson}: between nodes~$i,j$ belonging to groups~$r,s$ we place a number of edges which is Poisson distributed with mean~$\omega_{rs}$, except for self-edges $i=j$ for which the mean is~$\half\omega_{rr}$.  The edge frequencies~$\omega_{rs}$ thus dictate the relative probabilities of within- and between-group connections.  In the most widely studied case, the diagonal elements~$\omega_{rr}$ are chosen to be larger than the off-diagonal ones, so that edges are more likely within groups than between them, a type of structure known as assortative mixing or homophily.  Other types of structure are also possible, however, and are observed in some networks, such as disassortative structure in which between-group edges are more likely than in-group ones~\cite{Newman03c}.

Let us denote by $g_i$ the label of the group to which node~$i$ is assigned.  Then, given the parameters~$\mu_r$ and~$\omega_{rs}$, the probability of generating a complete set of group assignments~$\vec{g} = \set{g_i}$ and a network~$\mat{A}$ in this model is
\begin{align}
P(\vec{g},\mat{A}|\mu,\omega) &= \prod_i \mu_{g_i}
  \prod_{i<j} {\omega_{g_ig_j}^{A_{ij}}\over A_{ij}!}
  \exp\bigl(-\omega_{g_ig_j}\bigr) \nonumber\\
  &\qquad{}\times\prod_i {\bigl(\half\omega_{g_ig_i}\bigr)^{A_{ii}/2}\over
  \bigl(\half A_{ii}\bigr)!} \exp\bigl(-\half\omega_{g_ig_i}\bigr),
\label{eq:sbmlike}
\end{align}
where $\mu$ and $\omega$ now represent the complete sets of corresponding parameters.  The log of this probability is
\begin{align}
\log P(\vec{g},\mat{A}|\mu,\omega) &= \sum_i \log \mu_{g_i}
  + \half \sum_{ij} \bigl( A_{ij} \log \omega_{g_ig_j} - \omega_{g_ig_j} \bigr)
  \nonumber\\
  &\hspace{-4em}{}- \sum_{i<j} \log A_{ij}! - \sum_i \bigl[ \half A_{ii} \log 2
  + \log \bigl(\half A_{ii})! \bigr].
\label{eq:sbmll}
\end{align}
Again this separates terms that involve the parameters~$\mu$ and~$\omega$ from those that do not, making derivatives like those in Eq.~\eqref{eq:mstep1} simpler.

In the formalism considered in Section~\ref{sec:em}, the structure of the network is the only kind of unobserved data, but in the stochastic block model there are two kinds: the network~$\mat{A}$ and the group assignments~$\vec{g}$.  Our EM algorithm carries over straightforwardly to this case, but with the joint distribution of $\vec{g}$ and $\mat{A}$ taking the place of the distribution of~$\mat{A}$ alone.  When combined with a suitable data model, this approach allows us to infer both the network structure and the community structure from a single calculation.

An alternative approach, considered by Le~\etal~\cite{LLL18}, is to assume that the community structure is known via other means and only the network structure is to be inferred using the EM algorithm.  In that case, Eq.~\eqref{eq:sbmlike} is replaced with
\begin{align}
P(\mat{A}|\vec{g},\mu,\omega) &=
  \prod_{i<j} {\omega_{g_ig_j}^{A_{ij}}\over A_{ij}!}
  \exp\bigl(-\omega_{g_ig_j}\bigr) \nonumber\\
  &\qquad{}\times\prod_i {\bigl(\half\omega_{g_ig_i}\bigr)^{A_{ii}/2}\over
  \bigl(\half A_{ii}\bigr)!} \exp\bigl(-\half\omega_{g_ig_i}\bigr),
\end{align}
and $\vec{g}$ is treated as ``data'' with known values.  Then the EM algorithm once again gives a posterior distribution on the network structure alone.  In their work Le~\etal\ found the community structure using a traditional spectral algorithm.

\subsection{The configuration model and degree correction}
\label{sec:cm}
A potential issue with the models of the previous sections is that they all generate networks with Poisson degree distributions, which are quite unlike the strongly right-skewed degree distributions seen in many real-world networks~\cite{BA99b,ASBS00}.  We can circumvent this issue and create more realistic models using \defn{degree correction}.

The simplest example of a degree-corrected model is the configuration model, a random graph model that allows for arbitrary degree distributions~\cite{MR95,NSW01}.  In the configuration model one fixes the degree of each node separately and then places edges at random, but respecting the chosen node degrees.  The standard way to do this is to place ``stubs'' of edges at each node, equal in number to the chosen degree, then join pairs of stubs together at random to create complete edges.  It can be shown~\cite{Newman18c} that the number of edges falling between nodes~$i$ and~$j$ in such a network is Poisson distributed with mean $d_i d_j/\sum_k d_k$, where $d_i$ is the degree of node~$i$.

In our calculations we make use of a variant of the configuration model, similar to one proposed by Chung and Lu~\cite{CL02b}, in which, rather than employing edge stubs, one simply places a Poisson distributed number of edges between each pair of nodes with the appropriate mean.  We define a set of real-valued parameters~$\phi_i$, one for each node~$i$, then place a number of edges between each pair of nodes~$i,j$ which is Poisson distributed with mean $\omega\phi_i\phi_j$, or half that number if $i=j$.  Note that, like the model of Section~\ref{sec:poisson}, this model can produce networks with self-edges or multiedges (or both), which is somewhat unrealistic.  One commonly allows them nonetheless because it leads to technical simplifications and does not in practice make much difference in the common case of a sparse network.

As defined, the parameters~$\phi_i$ and $\omega$ are not identifiable: one can increase all $\phi_i$ by any constant factor without changing the model if one also decreases $\omega$ by the square of the same factor.  We can fix the values of the parameters by choosing a normalization for the~$\phi_i$.  This can be done in several ways, all of which ultimately give equivalent results, but for present purposes a convenient choice is to set the average of the $\phi_i$ equal to~1:
\begin{equation}
{1\over n} \sum_i \phi_i = 1.
\label{eq:sumphi}
\end{equation}
This choice has the nice feature that the average of the elements of the adjacency matrix is then given by
\begin{equation}
{1\over n^2} \biggl[ \sum_{i\ne j} \omega\phi_i\phi_j + 2 \sum_i \half\omega\phi_i^2 \biggr]
  = {\omega\over n^2} \sum_{ij} \phi_i\phi_j = \omega.
\end{equation}
Thus $\omega$ is the average value of an adjacency matrix element, just as in the earlier model of Section~\ref{sec:poisson}.

Given the parameters of the model, the probability~$P(\mat{A}|\phi,\omega)$ of generating a particular network is
\begin{align}
P(\mat{A}|\phi,\omega) &= \prod_{i<j} {(\omega\phi_i\phi_j)^{A_{ij}}\over A_{ij}!}
  \e^{-\omega\phi_i\phi_j} \nonumber\\
  &\hspace{3em}{}\times \prod_i {(\half\omega\phi_i^2)^{A_{ii}/2}\over
   \bigl(\half A_{ii}\bigr)!} \e^{-\omega\phi_i^2/2},
\label{eq:config}
\end{align}
and its log is
\begin{align}
\log P(\mat{A}|\phi,\omega)
  &= \half \sum_{ij} A_{ij} \log\omega + \sum_{ij} A_{ij} \log \phi_i
  - \half n^2\omega \nonumber\\
  &\hspace{-2em}{} - \sum_{i<j} \log A_{ij}!
  - \sum_i \bigl[ \half A_{ii} \log 2 + \log(\half A_{ii}\bigr)! \bigr],
\label{eq:cmlog}
\end{align}
where we have made use of Eq.~\eqref{eq:sumphi}.

One can apply the same degree correction approach to the stochastic block model of Section~\ref{sec:sbm}, which leads to the so-called degree-corrected stochastic block model~\cite{KN11a}.  In this model we again divide nodes among $k$ groups with probability~$\mu_r$ of assignment to group~$r$, but now between each pair of nodes~$i,j$ we place a number of edges that is Poisson distributed with mean~$\omega_{rs}\phi_i\phi_j$, where $r$ and $s$ are respectively the groups to which nodes $i$ and $j$ belong.  The additional factor of $\phi_i\phi_j$ allows us to control the degrees of the nodes and give the network essentially any degree distribution we desire.  We can fix the normalization of the parameters in various ways, for example by choosing the mean of~$\phi_i$ to be~1 within each individual group thus:
\begin{equation}
{1\over n_r} \sum_i \delta_{r,g_i} \phi_i = 1
\label{eq:dcsumphi}
\end{equation}
for all~$r$, with $n_r = \sum_i \delta_{r,g_i}$ being the number of nodes in group~$r$.

\section{Data models}
\label{sec:datamodels}
We now turn to data models, meaning models of the measurement process.  These models represent the way the data measured in our experiments depend on the under\-lying ground-truth network.

\subsection{Independent edge measurements}
\label{sec:indepedges}
Perhaps the simplest data model is one in which observations of edges are independent identically distributed Bernoulli random variables, conditioned only on the presence or absence of an edge in the same place in the ground-truth network.  That is, we make a measurement on a node pair~$i,j$ and it returns a simple yes-or-no answer about whether the nodes are connected by an edge, which depends only on the adjacency matrix element~$A_{ij}$ for the same node pair and any parameters of the process, and is independent of other matrix elements or any other measurements we may make.  That is not to say, however, that the answers we get need be accurate, and in general we will assume that they are not.  In an error-prone world, our measurements will sometimes reflect the truth about whether an edge exists and sometimes they will not.

Consider the simplest case in which $A_{ij}$ takes only the values zero and one.  We can then parametrize the possible outcomes of a measurement by two probabilities: the true-positive rate~$\alpha$, which is the probability of observing an edge where one truly exists, and the false-positive rate~$\beta$, which is the probability of observing an edge where none exists.  (The two remaining possibilities, of true negatives and false negatives, occur with probabilities $1-\beta$ and $1-\alpha$ respectively, so no additional parameters are needed to represent the rates of these events.)  The probability of observing an edge between nodes $i$ and$~j$ can then be succinctly written as $\alpha^{A_{ij}} \beta^{1-A_{ij}}$ and the probability of not doing so is $(1-\alpha)^{A_{ij}} (1-\beta)^{1-A_{ij}}$.

We give an application of this model to an example data set in Section~\ref{sec:reality}.

\subsection{Multiple edge types}
\label{sec:multitypes}
In Section~\ref{sec:types} we introduced a network model in which edges have several types, representing for instance different strengths of acquaintance in a social network.  The $k$ edge types were represented by integer values of adjacency matrix elements $A_{ij} = 0\ldots k-1$.  Observed data for such a network could take several forms.  For instance, one can imagine situations in which it might be possible, via a measurement of some kind, to determine not only whether an edge exists between two nodes but also what type of edge it is.  Such a situation could be represented by a set of variables that parametrize the probability of observing an edge of type~$j$ between a pair of nodes if there is an edge of type~$k$ in the ground truth.  This, however, leads to a rather complicated data model.  A simpler set-up is one in which measurements return only a yes-or-no answer about whether two nodes are connected by an edge and no information about edge type.  This can be represented by a model with separate parameters~$\alpha_0\ldots\alpha_{k-1}$ equal to the probability of observing an edge given each of the different ground-truth edge states.  Then the probability of observing an edge between nodes $i$ and~$j$ is simply~$\alpha_{A_{ij}}$ and the probability of not observing one is $1-\alpha_{A_{ij}}$.

\subsection{Multimodal data}
\label{sec:multimodal}
There are many cases where the data for a network consist not merely of one type of measurement but of two or more.  For instance a social network might be measured by surveying participants using traditional questionnaires or interviews, but also by collecting social media data, email or text messages, or using observations of face-to-face interactions~\cite{Aharony11,KN11c,Stopczynski14}.  A protein--protein interaction network might be measured using a combination of co-immunoprecipitation, affinity purification, yeast two-hybrid screens, or other methods~\cite{vonMering05}.  When represented as networks, such data are sometimes called multilayer or multiplex networks~\cite{Boccaletti14,DGPA16}.

Measurements of different types can be governed by different probabilities and errors.  Assuming a ground-truth network represented by a simple binary adjacency matrix with~$A_{ij}=0$ or~1, one could define separate true- and false-positive probabilities~$\alpha_m,\beta_m$ for each type of measurement.  That is, $\alpha_m$~is the probability that a measurement of type~$m$ will reveal an edge between two nodes~$i,j$ where an edge truly exists ($A_{ij}=1$), and $\beta_m$~is the probability that such a measurement will reveal an edge where none exists ($A_{ij}=0$).  Then the total probability of observing an edge between $i$ and~$j$ using a measurement of type~$m$ is $\alpha_m^{A_{ij}} \beta_m^{1-A_{ij}}$ and the probability of not observing one is $(1-\alpha_m)^{A_{ij}} (1-\beta_m)^{1-A_{ij}}$.

\subsection{Directed edges and individual node errors}
\label{sec:directed}
The models we have described so far assume undirected edges, but it is straightforward to generalize them to the case of directed networks.  Directed versions of the basic network models exist already, such as directed versions of the configuration model~\cite{NSW01} or the stochastic block model~\cite{WW87}.  Data models for directed networks are a natural generalization of the undirected versions.  For instance, one could assume that the empirical observations of interactions between nodes in a directed network are independent directed Bernoulli random variables that depend on the underlying ground-truth edges, with appropriately defined true- and false-positive rates.  In most cases the equations for the models are straightforward generalizations of those for the undirected case.  We give an example in Section~\ref{sec:multimodal2}.

In some cases it is possible for the observations of edges to be directed even if the underlying ground-truth network is undirected, or \textit{vice versa}.  Perhaps the most prominent example of this phenomenon arises in the study of social networks such as friendship or acquaintance networks.  In studies of these networks, by far the most common method for collecting data is simply to ask people who their friends or acquaintances are.  This results in directed edge measurements in which the fundamental unit of data is a statement by person $i$ that they are acquainted with person~$j$.  Often, however, we would consider the underlying network itself to be undirected---either two people are acquainted or they are not.  This situation can again be represented with a relatively straightforward generalization of earlier data models in which directed observations depend on the underlying undirected ground truth, with appropriately defined true- and false-positive rates.

Directed measurements like these give rise to the possibility that two people may make contradictory statements about whether they are acquainted: person~$i$ may claim to know person~$j$ but person~$j$ may claim not to know~$i$.  Such unreciprocated claims are in fact common in social network studies~\cite{VK08}.  In surveys of friendship among schoolchildren, for instance, only about a half of all claimed friendships are reciprocated~\cite{BN13}.  Such a situation can arise naturally in the data model: if the true- and false-positive rates for observations are $\alpha$ and $\beta$ as before, the probability of both of two individuals claiming acquaintance is $\alpha^{2A_{ij}} \beta^{2(1-A_{ij})}$, the probability of one but not the other doing so is $2[\alpha(1-\alpha)]^{A_{ij}} [\beta(1-\beta)]^{1-A_{ij}}$, and the probability of neither is $(1-\alpha)^{2A_{ij}} (1-\beta)^{2(1-A_{ij})}$.

An interesting alternative formulation, proposed by Butts~\cite{Butts03}, considers the case in which some individuals are more reliable in the reports they give than others.  Variations in reliability could arise simply because some people take surveys more seriously than others, are more cooperative survey subjects, or take more care with their responses.  But they could also arise because people have different perceptions of what it means to be acquainted: one person could have a relatively relaxed view in which they consider people to be acquaintances even if they barely know them, while another could adopt a stricter definition.

Such a situation can be represented by a data model in which there is a separate true-positive rate~$\alpha_i$ and false-positive rate~$\beta_i$ for each node~$i$.  Then the probability for instance of $i$ saying they are friends with $j$ but $j$ saying they are not is $[\alpha_i(1-\alpha_j)]^{A_{ij}} [\beta_i(1-\beta_j)]^{1-A_{ij}}$, and similar expressions apply for other patterns of observations.  Using this kind of model allows us to infer not only the structure of the underlying network but also the individual true- and false-positive rates, which themselves may reveal interesting behaviors---see Section~\ref{sec:individual}.

Surveys of social networks are not the only context in which directed measurements of undirected networks arise.  For instance, there have been many studies of messaging behavior within groups of people: who calls whom on the telephone, who emails whom, who sends text messages to whom, and so forth~\cite{EMB02,Onella07,LH08a,Stopczynski14}.  One can hypothesize that observations such as phone calls or emails are a noisy measurement of an underlying network of acquaintance, and hence use models such as those described above to infer the structure of the network from the observed pattern of messages.

\subsection{Networks with multiedges}
\label{sec:multidata}
Some ground-truth networks may be \defn{multigraphs}, meaning that they contain multiedges.  True multigraphs are rare in real-world applications, but there are many networks which, though composed of single edges only, may nonetheless be conveniently represented as multigraphs, for instance using the Poisson formulation of Section~\ref{sec:poisson}.  How should we define a data model for a multigraph?  There are a number of types of data a measurement of such a network could return.  It could, for instance, return an estimate of the multiplicity of an edge.  In our work, however, we make a simpler assumption, similar to that of Section~\ref{sec:multitypes}, in which measurements return only a yes-or-no answer that there either is or is not an edge at a given position.  This situation is most completely represented by a model with an infinite set of parameters~$\alpha_k$, representing the probability that upon making a measurement of a particular pair of nodes we observe an edge between them if there are exactly~$k$ edges in the corresponding position in the ground-truth network.  In practice, however, since multiedges are rare in the examples we consider, only the first two of these parameters are of interest: $\alpha_0$,~which is the false-positive rate, and~$\alpha_1$, which is roughly, though not exactly, the true-positive rate.

\section{Complete algorithms and example applications}
Building a complete algorithm for inferring network structure from noisy data involves combining a suitable network model with a suitable data model.  There are many such combinations we can construct from the models introduced in the previous sections.  Here we give a selection of examples, along with illustrative applications.

\subsection{Random graphs and independent measurements}
\label{sec:rgindep}
Perhaps the simplest example of our methods is the combination of the standard (Bernoulli) random graph of Section~\ref{sec:rg} with the independent edge data model of Section \ref{sec:indepedges}.  This turns the problem of network reconstruction into a standard binary classification problem.  We will go through this case in detail.

To derive the EM equations for this combination of models, we first take the probability~$P(\mat{A}|\omega)$ for the random graph from Eq.~\eqref{eq:rglike} and the uniform prior probability~$P(\omega)=1$ and substitute them into Eq.~\eqref{eq:mstep1}.  Performing the derivative with respect to the single parameter~$\omega$, we get
\begin{equation}
\sum_{\mat{A}} q(\mat{A}) \sum_{i<j} \biggl[ {A_{ij}\over\omega}
  - {1-A_{ij}\over1-\omega} \biggr] = 0.
\label{eq:rgderiv}
\end{equation}
Swapping the order of the summations and defining
\begin{equation}
Q_{ij} = \sum_{\mat{A}} q(\mat{A}) A_{ij},
\label{eq:defsq}
\end{equation}
we find that
\begin{equation}
\omega = {1\over{n\choose2}} \sum_{i<j} Q_{ij},
\label{eq:rgomega}
\end{equation}
where $n$ is the number of nodes in the network, as previously.

The quantity~$Q_{ij}$ is equal to the posterior probability that there is an edge between nodes $i$ and~$j$---it is our estimate of the ground truth for this node pair given the observed data.  $Q_{ij}$~can be thought of as a generalization of the adjacency matrix.  When it is exactly zero or one it has the same meaning as the adjacency matrix element~$A_{ij}$: there definitely either is or is not a ground-truth edge between nodes $i$ and~$j$.  For other values between zero and one it interpolates between these limits, quantifying our certainty about whether the edge exists.  Equation~\eqref{eq:rgomega} thus has the simple interpretation that the probability~$\omega$ of an edge in our network is the average of the probabilities of the individual edges.

Turning to the data model, a crucial point to notice is that if measurements of different edges are truly independent, so that an observation (or not) of an edge between one node pair tells you nothing about any other node pair, then single measurements of node pairs are not enough to estimate the parameters of the model.  It is well known that you cannot estimate the error on a random variable by making only a single measurement.  You have to make at least two measurements.  In the present context, this means that at least some edges in the network must be measured more than once to obtain an estimate of the true- and false-positive rates $\alpha$ and~$\beta$.

Let us assume that we make some number~$N_{ij}$ of measurements of node pair~$i,j$.  Each measurement returns a yes-or-no answer about whether the nodes are connected by an edge, but repeated measurements may not agree, precisely because the measurements are noisy.  So suppose that out of the $N_{ij}$ measurements we make, we observe an edge on $E_{ij}$ of them, and no edge on the remaining $N_{ij}-E_{ij}$.  Plugging these definitions into the data model of Section~\ref{sec:indepedges}, we can write the probability of this particular set of observations as $\alpha^{E_{ij}}(1-\alpha)^{N_{ij}-E_{ij}}$ if there is truly an edge between $i$ and $j$, and $\beta^{E_{ij}}(1-\beta)^{N_{ij}-E_{ij}}$ if there is not.  Taking the product over all distinct node pairs, the probability for the entire data set can then be written
\begin{align}
P(D|\mat{A},\alpha,\beta) &= \prod_{i<j}
  \bigl[ \alpha^{E_{ij}}(1-\alpha)^{N_{ij}-E_{ij}} \bigr]^{A_{ij}} \nonumber\\
  &\hspace{4em}{}\times\bigl[ \beta^{E_{ij}}(1-\beta)^{N_{ij}-E_{ij}}
                       \bigr]^{1-A_{ij}}.
\label{eq:abdata}
\end{align}
Taking the log and substituting into Eq.~\eqref{eq:mstep2}, assuming that the priors on $\alpha$ and $\beta$ are uniform, we get
\begin{align}
\sum_{\mat{A}} q(\mat{A}) \sum_{i<j} A_{ij} \biggl[ {E_{ij}\over\alpha}
  - {N_{ij}-E_{ij}\over1-\alpha} \biggr] &= 0, \\
\sum_{\mat{A}} q(\mat{A}) \sum_{i<j} (1-A_{ij}) \biggl[ {E_{ij}\over\beta}
  - {N_{ij}-E_{ij}\over1-\beta} \biggr] &= 0,
\end{align}
which can be rearranged to give
\begin{equation}
\alpha = {\sum_{i<j} Q_{ij} E_{ij}\over\sum_{i<j} Q_{ij} N_{ij}}, \qquad
\beta = {\sum_{i<j} (1-Q_{ij}) E_{ij}\over\sum_{i<j} (1-Q_{ij}) N_{ij}},
\label{eq:simpleab}
\end{equation}
where $Q_{ij}$ is as in Eq.~\eqref{eq:defsq} again.

\begin{widetext}
It remains to calculate the value of~$Q_{ij}$, which we do from Eq.~\eqref{eq:estep}.  Combining Eqs.~\eqref{eq:bayes}, \eqref{eq:rglike}, and~\eqref{eq:abdata} and substituting into~\eqref{eq:estep}, we find the following expression for~$q(\mat{A})$:
\begin{align}
q(\mat{A}) &= { \prod_{i<j} \bigl[ \omega \alpha^{E_{ij}}
  (1-\alpha)^{N_{ij}-E_{ij}} \bigr]^{A_{ij}}
  \bigl[ (1-\omega) \beta^{E_{ij}} (1-\beta)^{N_{ij}-E_{ij}} \bigr]^{1-A_{ij}}\over
  \sum_{\mat{A}} \prod_{i<j} \bigl[ \omega \alpha^{E_{ij}}
  (1-\alpha)^{N_{ij}-E_{ij}} \bigr]^{A_{ij}}
  \bigl[ (1-\omega) \beta^{E_{ij}} (1-\beta)^{N_{ij}-E_{ij}} \bigr]^{1-A_{ij}}}
  \nonumber\\
  &= \prod_{i<j} {\bigl[ \omega \alpha^{E_{ij}}
  (1-\alpha)^{N_{ij}-E_{ij}} \bigr]^{A_{ij}}
  \bigl[ (1-\omega) \beta^{E_{ij}} (1-\beta)^{N_{ij}-E_{ij}} \bigr]^{1-A_{ij}}\over
  \omega \alpha^{E_{ij}}
  (1-\alpha)^{N_{ij}-E_{ij}} + (1-\omega) \beta^{E_{ij}} (1-\beta)^{N_{ij}-E_{ij}}}.
\end{align}
Then
\begin{equation}
Q_{ij} = \sum_{\mat{A}} q(\mat{A}) A_{ij}
  = {\omega \alpha^{E_{ij}} (1-\alpha)^{N_{ij}-E_{ij}}\over
  \omega \alpha^{E_{ij}}
  (1-\alpha)^{N_{ij}-E_{ij}} + (1-\omega) \beta^{E_{ij}} (1-\beta)^{N_{ij}-E_{ij}}}.
\label{eq:simpleqij}
\end{equation}
\end{widetext}

The posterior distribution~$q(\mat{A})$ can be conveniently rewritten in terms of~$Q_{ij}$ as
\begin{equation}
q(\mat{A}) = \prod_{i<j} Q_{ij}^{A_{ij}} (1-Q_{ij})^{1-A_{ij}}.
\label{eq:simplepost}
\end{equation}
In other words, the probability distribution over networks is (in this special case) simply the product of independent Bernoulli distributions of the individual edges, with Bernoulli parameters~$Q_{ij}$.

The complete EM algorithm now consists of the iteration of Eqs.~\eqref{eq:rgomega}, \eqref{eq:simpleab}, and~\eqref{eq:simpleqij} from suitably chosen starting conditions until convergence.  Typically one chooses random values of $\omega$, $\alpha$, and $\beta$ for the initial conditions and proceeds from there.

Once the algorithm has converged we can estimate network quantities of interest using Eqs.~\eqref{eq:mux} and~\eqref{eq:sigmax}.  As a simple example, consider the average degree~$c$ of a node in the network.  For a known network with adjacency matrix~$\mat{A}$ the average degree is given by $c = (1/n) \sum_{ij} A_{ij}$.  The mean (expected) value of the average degree given our posterior distribution~$q(\mat{A})$ is thus
\begin{align}
\mu_c &= \sum_{\mat{A}} q(\mat{A}) {1\over n} \sum_{ij} A_{ij}
       = {1\over n} \sum_{ij} \sum_{\mat{A}} q(\mat{A}) A_{ij} \nonumber\\
      &= {1\over n} \sum_{ij} Q_{ij}.
\label{eq:muc}
\end{align}
The estimated variance about this value is given by Eq.~\eqref{eq:sigmax} to be
\begin{align}
\sigma_c^2 &= \sum_{\mat{A}} q(\mat{A}) \biggl[ {1\over n} \sum_{ij} A_{ij}
              - \mu_c \biggl]^2 \nonumber\\
           &= {1\over n^2} \sum_{\mat{A}} q(\mat{A}) \sum_{ijkl} A_{ij} A_{kl}
              - \mu_c^2 \nonumber\\
           &= {1\over n^2} \sum_{ij} Q_{ij}(1-Q_{ij}).
\label{eq:sigmac}
\end{align}

The approach of this section generalizes straightforwardly to the variant random graph of Section~\ref{sec:poisson} in which there is a Poisson distributed number of edges between each pair of nodes and the network can contain multiedges.  Taking Eq.~\eqref{eq:logpoisson} and substituting into~\eqref{eq:mstep1} we get
\begin{equation}
\sum_{\mat{A}} q(\mat{A}) \sum_{ij} \biggl[ {A_{ij}\over\omega} - 1 \biggr]
  = 0.
\label{eq:poissonmstep}
\end{equation}
Here we have again assumed a uniform prior on~$\omega$, which is not strictly allowed in this case, since $\omega$ has an infinite range from 0 to~$\infty$.  One can, however, assume a uniform prior over a finite range and then make that range large enough to encompass the solution for~$\omega$.

Rearranging Eq.~\eqref{eq:poissonmstep} for~$\omega$ now gives
\begin{align}
\omega &= {1\over n^2} \sum_{\mat{A}} q(\mat{A}) \sum_{ij} A_{ij}
   = {1\over n^2} \sum_{ij} \sum_{\mat{A}} q(\mat{A})
     \sum_{k=0}^\infty k\,\delta_{k,A_{ij}} \nonumber\\
  &= {1\over n^2} \sum_{ij} \sum_{k=0}^\infty k Q_{ij}(k),
\label{eq:poissonomega}
\end{align}
where
\begin{equation}
Q_{ij}(k) = \sum_{\mat{A}} q(\mat{A})\,\delta_{k,A_{ij}}
\label{eq:qijk}
\end{equation}
is the posterior probability that there are exactly~$k$ edges between nodes $i$ and~$j$ (or $\half k$ edges when $i=j$).  Alternatively, and perhaps more conveniently, we can write the estimated value of $A_{ij}$ as
\begin{equation}
\hat A_{ij} = \sum_{k=0}^\infty k Q_{ij}(k),
\label{eq:ahat}
\end{equation}
in which case
\begin{equation}
\omega = {1\over n^2} \sum_{ij} \hat A_{ij}.
\label{eq:omegaahat}
\end{equation}

As discussed in Section~\ref{sec:multidata}, we will assume that, multiedges notwithstanding, measurements on node pairs~$i,j$ continue to return yes-or-no answers about the presence of an edge, with $\alpha_k$ being the probability of a yes if there are $k$ ground-truth edges between $i$ and~$j$.  Let $E_{ij}$~represent the number of yeses out of a total of~$N_{ij}$ measurements, except for self-edges, for which the most natural definition is that $E_{ii}$ represents twice the number of yeses and $N_{ii}$ twice the number of measurements, by analogy with the definition of the adjacency matrix.

With these definitions, the equivalent of Eq.~\eqref{eq:abdata} for this model is
\begin{align}
P(D|\mat{A},\alpha) &= \prod_{i<j} \, \prod_{k=0}^\infty
  \bigl[ \alpha_k^{E_{ij}}(1-\alpha_k)^{N_{ij}-E_{ij}} \bigr]^{\delta_{k,A_{ij}}}
  \nonumber\\
  &\hspace{-1em}{}\times
  \prod_i \, \prod_{k=0}^\infty
  \bigl[ \alpha_k^{E_{ii}/2}(1-\alpha_k)^{(N_{ii}-E_{ii})/2}
  \bigr]^{\delta_{k,A_{ii}}}.
\label{eq:poissondata}
\end{align}
Taking the log, substituting into Eq.~\eqref{eq:mstep2}, and assuming that the priors on the $\alpha_k$ are uniform, we then get
\begin{equation}
\sum_{\mat{A}} q(\mat{A}) \sum_{ij} \delta_{k,A_{ij}}
  \biggl[ {E_{ij}\over\alpha_k} - {N_{ij}-E_{ij}\over1-\alpha_k} \biggr] = 0
\end{equation}
for all~$k=0\ldots\infty$.  Rearranging for~$\alpha_k$, we get
\begin{equation}
\alpha_k = {\sum_{ij} Q_{ij}(k) E_{ij}\over\sum_{ij} Q_{ij}(k) N_{ij}},
\label{eq:poissonalphak}
\end{equation}
where $Q_{ij}(k)$ is defined in Eq.~\eqref{eq:qijk}.

\begin{widetext}
Following similar lines of argument to those for the Bernoulli model, Eq.~\eqref{eq:estep} now tells us that the posterior distribution over networks~$\mat{A}$ is
\begin{equation}
q(\mat{A}) = \prod_{i<j} {\omega^{A_{ij}}/A_{ij}!
  \bigl[ \alpha_{A_{ij}}^{E_{ij}}(1-\alpha_{A_{ij}})^{N_{ij}-E_{ij}} \bigr]\over
  \sum_{k=0}^\infty \omega^k/k!
  \bigl[ \alpha_k^{E_{ij}}(1-\alpha_k)^{N_{ij}-E_{ij}} \bigr]}
  \prod_i {(\half\omega)^{A_{ii}/2}/(\half A_{ii})!
  \bigl[ \alpha_{A_{ii}}^{E_{ii}/2}(1-\alpha_{A_{ii}})^{(N_{ii}-E_{ii})/2} \bigr]
  \over \sum_{r=0}^\infty (\half\omega)^r/r!
  \bigl[ \alpha_{2r}^{E_{ii}/2}(1-\alpha_{2r})^{(N_{ii}-E_{ii})/2} \bigr]}
  = \prod_{i\le j} Q_{ij}(A_{ij}).
\label{eq:poissonpost}
\end{equation}
\end{widetext}
Then
\begin{equation}
Q_{ij}(k) = {\omega^k/k!
  \bigl[ \alpha_k^{E_{ij}}(1-\alpha_k)^{N_{ij}-E_{ij}} \bigr]\over
  \sum_{k=0}^\infty \omega^k/k!
  \bigl[ \alpha_k^{E_{ij}}(1-\alpha_k)^{N_{ij}-E_{ij}} \bigr]}
\label{eq:qijkfull}
\end{equation}
for $i\ne j$ and
\begin{equation}
Q_{ii}(k) = {(\half\omega)^{k/2}/(\half k)!
  \bigl[ \alpha_k^{E_{ii}/2}(1-\alpha_k)^{(N_{ii}-E_{ii})/2} \bigr]
  \over \sum_{r=0}^\infty (\half\omega)^r/r!
  \bigl[ \alpha_{2r}^{E_{ii}/2}(1-\alpha_{2r})^{(N_{ii}-E_{ii})/2} \bigr]}.
\end{equation}
In the common case of network that does not actually have any self-edges, however, one would not normally attempt to measure their presence, so $N_{ii}=E_{ii}=0$ for all~$i$ and the latter expression simplifies to
\begin{equation}
Q_{ii}(k) = {(\half\omega)^{k/2}\over(\half k)!} \e^{-\omega/2},
\label{eq:qiikfull}
\end{equation}
which is simply the prior distribution on self-edges assuming the random graph model.  In practice, for sparse networks where $\omega$ is small, it will often be an adequate approximation to simply set $Q_{ii}(0)=1$ for all~$i$ and $Q_{ii}(k)=0$ for $k>0$, implying that there are no self-edges, which is true.

In theory, the evaluation of $Q_{ij}(k)$ from Eq.~\eqref{eq:qijkfull} requires us to first calculate all of the parameters~$\alpha_k$, of which there are an infinite number, in order to evaluate the denominator.  In practice, however, most networks, as we have said, are sparse, having small values of~$\omega$, which means that all but the first two terms in the denominator can be neglected and only $\alpha_0$ and~$\alpha_1$ need be calculated (which represent approximately the false-positive and true-positive rates for this data model).  This in turn means that $Q_{ij}(k)$ is negligible for $k\ge2$, so that $Q_{ij}(0) \simeq 1 - Q_{ij}(1)$.  Thus we only really need to calculate one probability~$Q_{ij}(1)$ for each node pair:
\begin{equation}
Q_{ij}(1) \simeq {\omega\alpha_1^{E_{ij}}(1-\alpha_1)^{N_{ij}-E_{ij}} \over
  \alpha_0^{E_{ij}}(1-\alpha_0)^{N_{ij}-E_{ij}}
  + \omega\alpha_1^{E_{ij}}(1-\alpha_1)^{N_{ij}-E_{ij}}},
\label{eq:poissonq}
\end{equation}
which represents, roughly speaking, the probability that there is an edge between~$i$ and~$j$, which is also (approximately) the expected value of the corresponding adjacency matrix element $\hat A_{ij} \simeq Q_{ij}(1)$.

\subsection{Configuration model with independent measurements}
\label{sec:cmindep}
The developments of the previous section can be extended in a straightforward manner to the more realistic configuration model introduced in Section~\ref{sec:cm}.  Substituting Eq.~\eqref{eq:cmlog} into Eq.~\eqref{eq:mstep1} and differentiating with respect to~$\omega$ gives
\begin{align}
\omega &= {1\over n^2} \sum_{\mat{A}} q(\mat{A}) \sum_{ij} A_{ij}
        = {1\over n^2} \sum_{ij} \sum_{k=0}^\infty k Q_{ij}(k) \nonumber\\
       &= {1\over n^2} \sum_{ij} \hat A_{ij},
\label{eq:cmomega}
\end{align}
just as in Eqs.~\eqref{eq:poissonomega} and~\eqref{eq:omegaahat}, with $Q_{ij}(k) = \sum_{\mat{A}} q(\mat{A}) \delta_{k,A_{ij}}$ as before and $\hat A_{ij} = \sum_k k Q_{ij}(k)$ being the estimated value of~$A_{ij}$, Eq.~\eqref{eq:ahat}.  At the same time, differentiating with respect to~$\phi_i$, while enforcing the normalization condition~\eqref{eq:sumphi} with a Lagrange multiplier, gives
\begin{equation}
\phi_i = n {\sum_j \hat A_{ij}\over\sum_{ij} \hat A_{ij}}.
\label{eq:cmphi}
\end{equation}

The equations for the data model parameters~$\alpha_k$ are unchanged from the previous section, with $\alpha_k$ still being given by Eq.~\eqref{eq:poissonalphak}.  And the posterior distribution over networks is once again given by $q(\mat{A})=\prod_{i\le j} Q_{ij}(A_{ij})$ but now with
\begin{equation}
Q_{ij}(k) = {(\omega\phi_i\phi_j)^k/k!
  \bigl[ \alpha_k^{E_{ij}}(1-\alpha_k)^{N_{ij}-E_{ij}} \bigr]\over
  \sum_{k=0}^\infty (\omega\phi_i\phi_j)^k/k!
  \bigl[ \alpha_k^{E_{ij}}(1-\alpha_k)^{N_{ij}-E_{ij}} \bigr]},
\label{eq:qijkcm}
\end{equation}
and $Q_{ii}(k) = \e^{-\omega\phi_i^2/2} (\half\omega\phi_i^2)^{k/2}/(\half k)!$.

If we make the same assumption as we made at the end of the previous section, that $\omega$ is small and hence only the first two terms in the denominator of Eq.~\eqref{eq:qijkcm} need be included, then one need only calculate the quantities
\begin{equation}
Q_{ij}(1) \simeq {\omega\phi_i\phi_j
  \alpha_1^{E_{ij}}(1-\alpha_1)^{N_{ij}-E_{ij}} \over
  \alpha_0^{E_{ij}}(1-\alpha_0)^{N_{ij}-E_{ij}}
  + \omega\phi_i\phi_j \alpha_1^{E_{ij}}(1-\alpha_1)^{N_{ij}-E_{ij}}},
\label{eq:cmq}
\end{equation}
and $Q_{ij}(0) \simeq 1 - Q_{ij}(1)$, $\hat A_{ij} \simeq Q_{ij}(1)$ (and $Q_{ii}(0)\simeq1$).

\begin{figure*}
\begin{center}
\hfill\subfigure[]{\includegraphics[width=8cm]{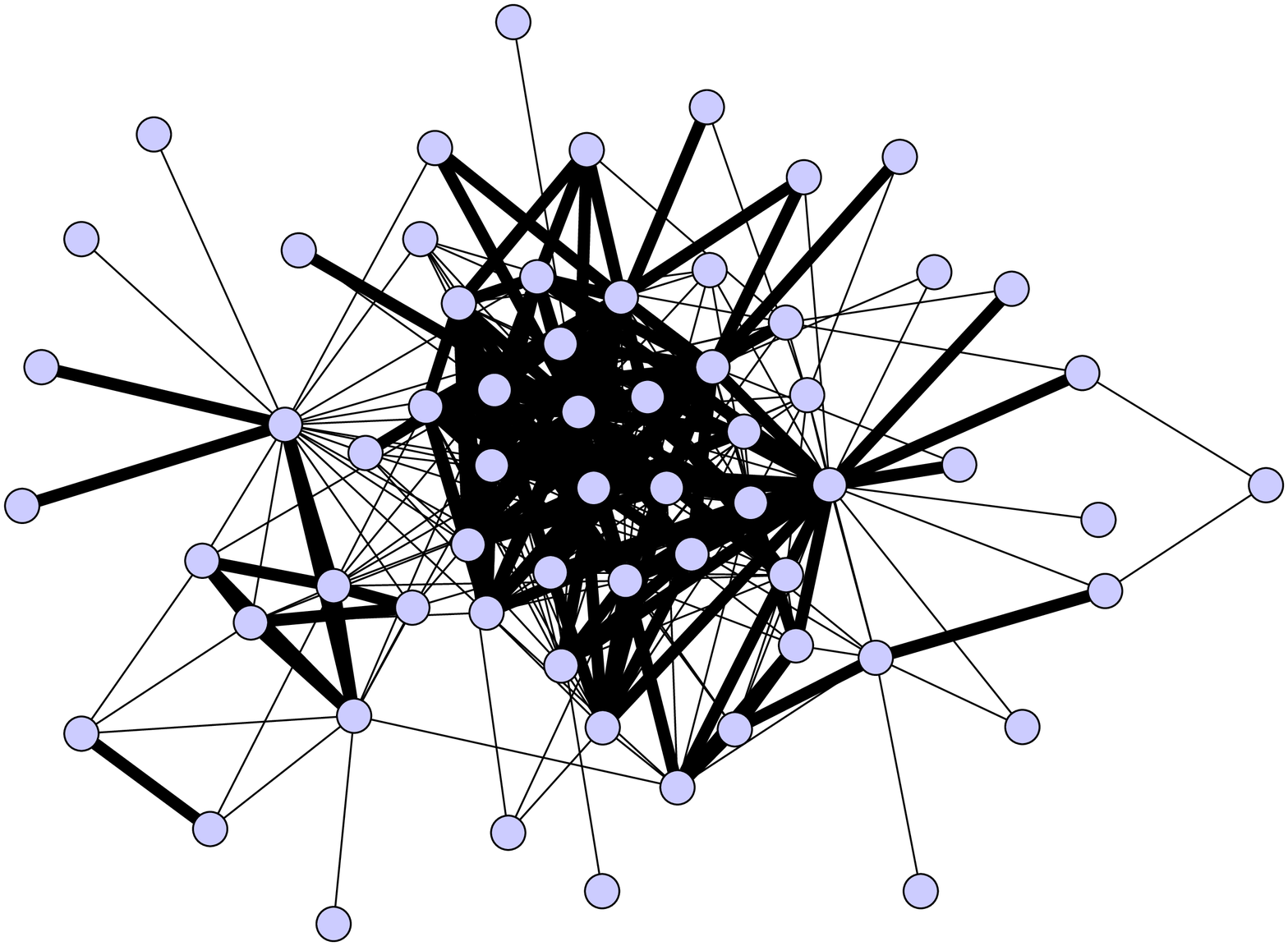}}\hspace{3em}
\subfigure[]{\includegraphics[width=8cm]{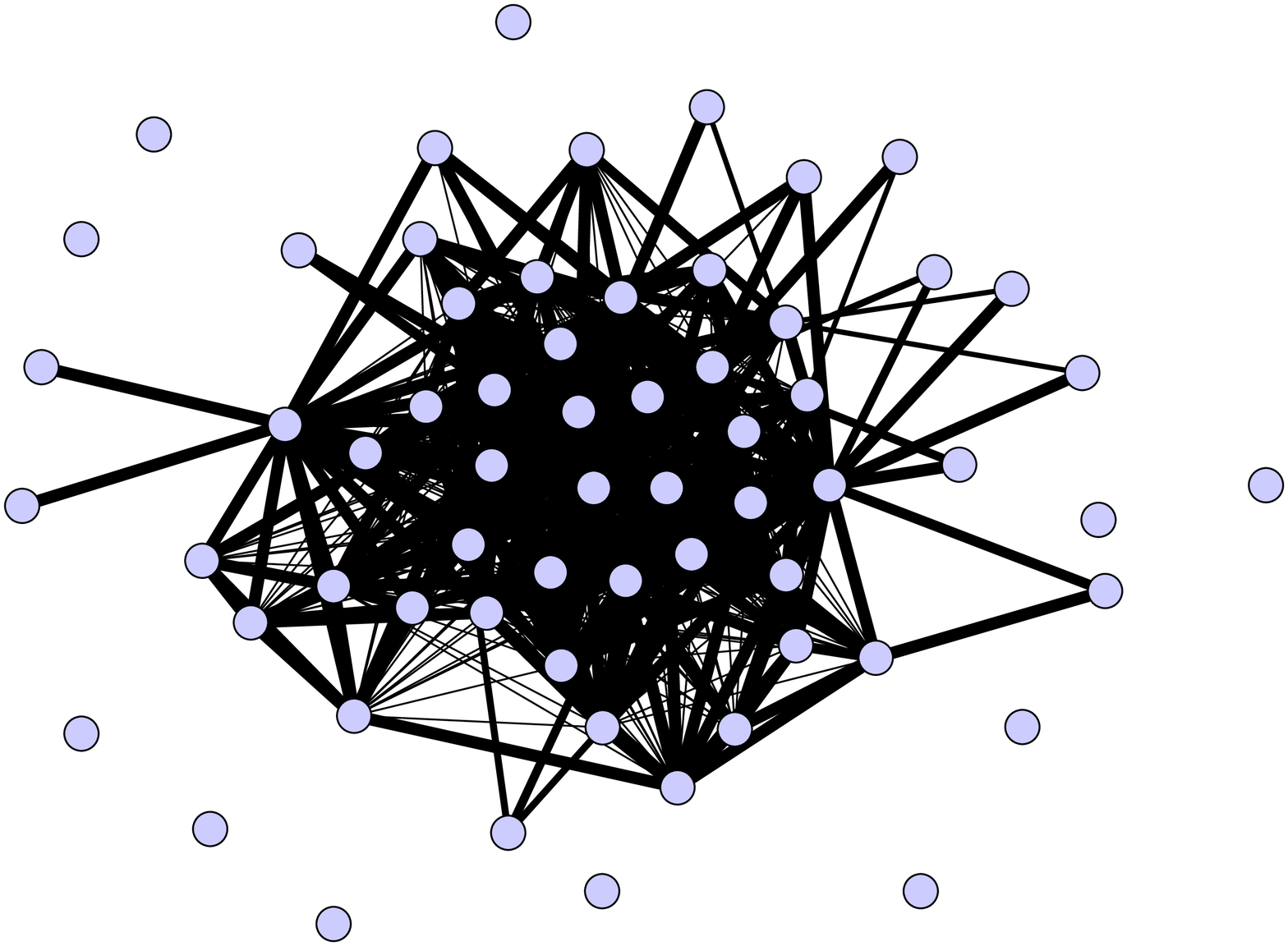}}\hfill\null
\end{center}
\caption{Two examples of inferred networks of connections between a subset of the participants in the ``reality mining'' study of Eagle and Pentland~\cite{EP06,EPL09}.  (a)~Network inferred using the Poisson random graph for the network model and the independent edges model of Section~\ref{sec:indepedges} for the data model.  Edges with probability less than 0.01 are omitted, as are nodes that have no edges with probability 0.01 or greater.  (b)~Network inferred from the same data, but using the configuration model as the network model.  For ease of comparison, the same set of nodes is shown in panel~(b) as in panel~(a), with each in the same spatial position.}
\label{fig:poissoncm}
\end{figure*}

\subsection{Example application}
\label{sec:reality}
As an example of the application of these methods, we turn to a data set we examined previously in~\cite{Newman18a}.  The data come from the ``reality mining'' study of Eagle and Pentland~\cite{EP06,EPL09} and describe the interactions of a group of 96 university students.  The goal of the study was to determine whether one could reconstruct networks of acquaintance---who actually knows whom---from data on physical proximity.  Students in the study carried mobile phones equipped with special software that used Bluetooth radio technology to record when two of the phones where in close proximity with one another (a few meters).  It is reasonable to suppose that people who are acquainted will sometimes be in close proximity, but it is also certainly the case that some acquaintances are rarely or never in proximity and that people may be in proximity and not be acquainted.  You might sit next to someone on the bus, for example, or stand next to them in the line at the supermarket without ever knowing who they are.  Thus proximity is a noisy measurement of acquaintance of exactly the kind considered here.

The study by Eagle and Pentland recorded detailed, time-resolved instances of pairwise proximity between participants over a period of several months in 2004 and 2005, but the data we study cover a smaller interval, being taken from eight consecutive Wednesdays in March and April of 2005.  We limit ourselves to Wednesdays in order to factor out the (large) weekly variation in proximity patterns: lower rates of proximity are observed at weekends than on weekdays for instance.  We also amalgamate all observations for each Wednesday into a single measurement: we consider two people to be observed together on a particular day if they are measured to be in proximity at any time during that day.  The result is eight separate measurements of proximity for each pair of individuals.  In the nomenclature of our models, $N_{ij} = 8$ for all $i,j$, and $E_{ij}$ can take integer values from 0 up to~8.  All possible values in this range are observed in the data.

Figure~\ref{fig:poissoncm}a shows what happens when we apply the algorithm of Eqs.~\eqref{eq:poissonalphak} and~\eqref{eq:poissonq} to these data.  This algorithm assumes a simple random graph for the network model and (conditionally) independent edge measurements for the data model.  The figure shows the resulting inferred network with edge thicknesses representing the posterior probabilities of the edges.  As we can see there is a well connected core of about twenty nodes in the center of the picture, surrounded by a periphery with weaker connections.  The thickest lines in the figure represent edges with probability of almost~1, while the thinnest represent edges with probability less than~0.1.  Edges with probability less than 0.01 are omitted from the figure, as are nodes that have no connections above this threshold.

Figure~\ref{fig:poissoncm}b shows the same data analyzed using the algorithm of Eqs.~\eqref{eq:cmomega}, \eqref{eq:cmphi}, and~\eqref{eq:cmq}, which uses the configuration model as its network model.  As we can see, this produces some changes in the inferred edge probabilities.  Overall many of the same edges get high or low probability in both models but the configuration-model based algorithm gives a more ``decisive'' result than its random-graph counterpart, mostly assigning either very high or very low probabilities to edges, meaning that it is more certain whether edges do or do not exist.

One way to think about the configuration model in this context is that it introduces correlations between edges that are not present when one uses the simple random graph.  The presence of edges attached to a particular node~$i$ increases the inferred value of the node parameter~$\phi_i$ via Eq.~\eqref{eq:cmphi}, and this in turn increases the probability of other edges being attached to the same the node via Eq.~\eqref{eq:cmq}.  Thus the presence of one edge makes the presence of another more likely.

One might ask which is the better of the two network models: the random graph or the configuration model?  In fact, despite the visual differences between the networks in Fig.~\ref{fig:poissoncm}, the two do not differ very greatly.  If we take the maximum-probability structure predicted by each algorithm (which is equivalent to assuming an edge to exist whenever $Q_{ij}>\half$ and not otherwise), then we find that the two calculations agree on essentially all node pairs \emph{except} those for which $E_{ij}=1$, i.e.,~those for which proximity was observed on only one of the eight days of observation.  For these pairs the random-graph-based calculation always concludes that the corresponding edge does not exist, whereas the configuration model sometimes says it does and sometimes says it doesn't.  Thus the primary contribution of the configuration model in this instance is to give us more nuanced results in the case of node pairs with particularly sparse observations.

More generally, the configuration model is considered to be the more accurate model in most real-world cases since it allows for realistic non-Poisson degree distributions similar to those seen in empirical networks~\cite{NSW01}.  Nonetheless, there maybe cases where the ordinary random graph is justified; which model one uses depends in the end on the assumptions one makes about the nature of the network.  One might perhaps consider this a problem with the method.  Other network reconstruction methods do not require one to make assumptions in this way.  As discussed at the start of Section~\ref{sec:em}, however, we would argue that these other methods are still making assumptions, though they may not be explicitly acknowledged.  We feel that the approach we propose is preferable in that it requires us to make our assumptions clear and allows us to see directly what effect they have on the results.

Note also that our algorithms cannot make any statement about what network it is exactly that is represented in pictures like Fig.~\ref{fig:poissoncm}.  Is it a network of who is friends with whom?  Who knows whom?  Who works with whom?  The algorithm does not say.  All we can say is that this is our best estimate of whatever network it is that is driving the observations.  In the present case it is probably some amalgam of friendship, students who work together, students who go to class together, and so forth.  ``Acquaintance'' might be a good umbrella term for this set of interactions, but in the end the network is most correctly defined as that network which causes people to be in proximity with one another.

Once we have the posterior distribution over networks, we can estimate any other network quantity of interest from it using Eqs.~\eqref{eq:mux} and~\eqref{eq:sigmax}.  For instance, we can calculate the mean degree~$c$ of the network, Eqs.~\eqref{eq:muc} and~\eqref{eq:sigmac}.  Using the configuration model version of our calculation and approximating $Q_{ij}$ by $Q_{ij}(1)$ we find that $c = 5.55 \pm 0.05$.  By contrast, a naive estimate of the mean degree, derived by simply aggregating all proximity observations for the eight days of data, gives $c = 6.23$, which is of the same order of magnitude, but nonetheless in significant disagreement (and lacking any error estimate).  A more conservative estimate, in which we assume an edge only if there are proximity observations between a given pair of nodes on two or more days, gives a lower value of $c = 3.00$, again in substantial disagreement with our estimate from the posterior distribution.

Running time for the calculation is minimal, varying from a fraction of a second to a few seconds depending on model details and the programming language used.  More generally, since the calculation requires the evaluation of $\Ord(n^2)$ probabilities~$Q_{ij}$, we expect the running time to scale at least as~$n^2$.  Network algorithms running in $\Ord(n^2)$ time are typically feasible (with patience) for networks of up to hundreds of thousands or perhaps millions of nodes, putting our methods within reach for many large network data sets, though not, in their current form, for the very largest (there exist examples with billions of nodes or more).

\subsection{Multimodal data}
\label{sec:multimodal2}
For our second example we consider the case discussed in Section~\ref{sec:multimodal} of a network whose edges are observed using a number of different methods or modes, labeled by $m=1\dots M$.  We will consider specifically a directed network, both to give an explicit example of an algorithm for directed edges and because it will be useful in Section~\ref{sec:foodweb}, where we will apply the method to a directed data set.  By convention, directed networks are represented by an adjacency matrix in which $A_{ij}=1$ if there is an edge \emph{from} node~$j$ \emph{to} node~$i$.

We will assume that the prior probability of any directed edge is~$\omega$ as previously.  Then, assuming a simple network in which there are no multiedges or self-edges, we have
\begin{equation}
P(\mat{A}|\omega) = \prod_{i\ne j} \omega^{A_{ij}} (1-\omega)^{1-A_{ij}},
\end{equation}
which is a trivial generalization of Eq.~\eqref{eq:rglike}.  Following the same line of argument as in Eq.~\eqref{eq:rgderiv} we can then show that
\begin{equation}
\omega = {1\over n(n-1)} \sum_{i\ne j} Q_{ij},
\label{eq:diromega}
\end{equation}
where $Q_{ij} = \sum_{\mat{A}} q(\mat{A}) A_{ij}$ is the posterior probability of a directed edge from node~$j$ to node~$i$.

Now let the true- and false-positive rates for observations in mode~$m$ be $\alpha_m$ and~$\beta_m$ respectively, as described in Section~\ref{sec:multimodal}.  And let $N_{ij}^{(m)}$ be the number of measurements made of the presence or absence of an edge from $j$ to $i$ (usually zero or one, but other values are possible in principle) and $E_{ij}^{(m)}$ be the number of times an edge is in fact observed.  Then the likelihood of the data~$D$ given the ground-truth network and parameters is
\begin{align}
P(D|\mat{A},\alpha,\beta) &= \prod_{i\ne j}
  \biggl[ \prod_{m=1}^M \alpha_m^{E_{ij}^{(m)}}
  (1-\alpha_m)^{N_{ij}^{(m)}-E_{ij}^{(m)}} \biggr]^{A_{ij}} \nonumber\\
  &\hspace{1em}{}\times\biggl[ \prod_{m=1}^M \beta_m^{E_{ij}^{(m)}}
  (1-\beta_m)^{N_{ij}^{(m)}-E_{ij}^{(m)}} \biggr]^{1-A_{ij}}.
\label{eq:abdata2}
\end{align}
Here we are assuming that measurements made in different modes are statistically independent, so that the probability of observing any given edge in any given combination of modes is a product over the probabilities of the individual modes.  In the language of machine learning such an approach is called a naive Bayes classifier.

\begin{widetext}
Following the same line of argument as in Eq.~\eqref{eq:abdata}, we can then show that
\begin{equation}
\alpha_m = {\sum_{i\ne j} Q_{ij} E_{ij}^{(m)}\over
           \sum_{i\ne j} Q_{ij} N_{ij}^{(m)}}, 
\qquad
\beta_m = {\sum_{i\ne j} (1-Q_{ij}) E_{ij}^{(m)}\over
           \sum_{i\ne j} (1-Q_{ij}) N_{ij}^{(m)}},
\label{eq:multiab}
\end{equation}
while the equivalent of Eq.~\eqref{eq:simpleqij} for~$Q_{ij}$ is
\begin{equation}
Q_{ij} = {\omega \prod_m \alpha_m^{E_{ij}^{(m)}}
         (1-\alpha_m)^{N_{ij}^{(m)}-E_{ij}^{(m)}}\over
         \omega \prod_m \alpha_m^{E_{ij}^{(m)}}
         (1-\alpha_m)^{N_{ij}^{(m)}-E_{ij}^{(m)}}
         + (1-\omega) \prod_m \beta_m^{E_{ij}^{(m)}}
         (1-\beta_m)^{N_{ij}^{(m)}-E_{ij}^{(m)}}}.
\label{eq:multiqij}
\end{equation}
The EM algorithm now consists of the iteration of Eqs.~\eqref{eq:diromega}, \eqref{eq:multiab}, and~\eqref{eq:multiqij} from suitable starting values to convergence.\end{widetext}

\subsection{Example application}
\label{sec:foodweb}
As an example of this algorithm, we consider an ecological network, a food web of predator-prey interactions between species.  The specific example we look at is the early Eocene Messel Shale food web of Dunne~\etal~\cite{DLW14}, a prehistoric food web of exactly $n=700$ extinct taxa and their patterns of predation, reconstructed from paleonto\-logical evidence.  Like many food webs, this one is pieced together from data derived from a variety of sources.  In this case, the authors used ten different types of evidence to establish links between taxa, including such things as gut contents (the digested remains of one species were found in the fossilized gut of another), stratigraphic co-occurrence (evidence of interaction is present in one or more other nearby contemporaneous fossil deposits), or body size (larger animals eat smaller ones, so a difference in body sizes can suggest a predator-prey interaction).

What is particularly interesting about this data set for our purposes is that Dunne~\etal\ made available not only the final form of the network but the details of which particular modes were observed for each edge in the network---gut contents, body size, etc.  Thus the data set has exactly the ``multimodal'' form we considered in the previous section.

Of the ten modes of observation used by Dunne~\etal\, one of them---``taxonomic uniformity''---is seen in virtually all edges (6126 edges out of a total of 6444, or 95\%), which means in practice that it communicates almost no information.  So we discard it, leaving $M=9$ remaining modes of measurement in the data set.  Each mode is listed as either observed or not for each edge in the network, meaning in effect that $N_{ij}^{(m)}=1$ for all $i,j$ and all~$m$, and $E_{ij}^{(m)}$ is either zero or one.  Applying Eqs.~\eqref{eq:diromega}, \eqref{eq:multiab}, and~\eqref{eq:multiqij} to these data and iterating to convergence, we then arrive at values for the true- and false-positive rates in each mode and probabilities~$Q_{ij}$ for the directed edges.

In addition to the data set itself, Dunne~\etal\ published their own judgments about the structure of the network.  For each edge that was observed in at least one of the modes they assigned a score, based on the data, indicating how confident they were that the edge in question actually exists in the network.  They used a three-valued scale to say whether they judged there to be high, medium, or low certainty about each edge.  If we similarly divide the edges of our inferred network into three categories, arbitrarily defining high certainty to be $Q_{ij}>0.9$, low certainty to be $Q_{ij}<0.1$, and medium certainty to be everything else, we find that our EM algorithm is able to reproduce the assessments of Dunne~\etal\ for 5446 of the 6444 observed edges, or 84.5\%.  For comparison, a random guess would get only 33\% correct.

In addition to the network itself, the inferred values of the parameters~$\alpha_m$ and~$\beta_m$ are also of interest.  Because this network (like most others studied in network science) is very sparse, all the~$\beta_m$ are small.  To make them easier to interpret we reparametrize them in terms of the \defn{false-discovery rate}, which is the probability that an observation of an edge is wrong.  Applying Bayes' rule, the false-discovery rate for observations in mode~$m$ is given by
\begin{align}
P(A_{ij}^{(m)}=0|&E_{ij}^{(m)}=1) \nonumber\\
  &= P(E_{ij}^{(m)}=1|A_{ij}^{(m)}=0)
  {P(A_{ij}^{(m)}=0)\over P(E_{ij}^{(m)}=1)} \nonumber\\
  &= {(1-\omega)\beta_m\over\omega\alpha_m + (1-\omega)\beta_m}.
\end{align}

\begin{figure}
\begin{center}
\includegraphics[width=8cm]{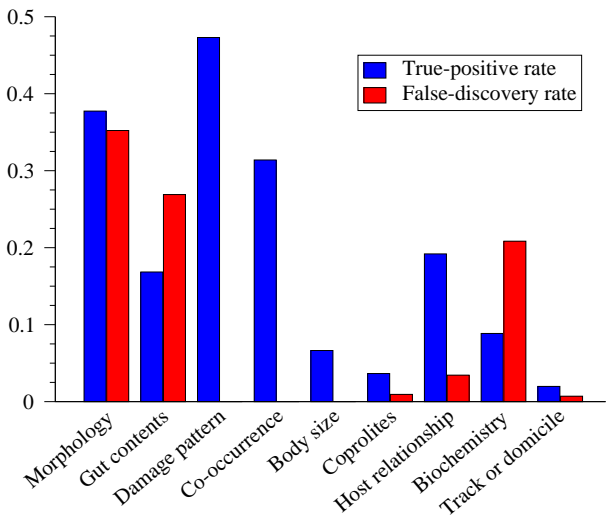}
\end{center}
\caption{Inferred values of the true-positive and false-discovery rates for each of the nine types of evidence considered in our analysis of the Messel Shale food web data set of Dunne~\etal~\cite{DLW14}.  High true-positive rate indicates a type of evidence that is frequently observed when there truly is a link in the network; high false-discovery rate indicates that many observations of links using this type of evidence are in error.}
\label{fig:messelab}
\end{figure}

Figure~\ref{fig:messelab} shows the estimated true-positive and false-discovery rates for each of the nine measurement modes.  A high true-positive rate for a mode means that when an edge truly exists we will typically see evidence in this mode.  A high false-discovery rate means that observations in this mode cannot be trusted because they are frequently false alarms.

The figure reveals that none of the modes of observation has a particularly high estimated true-positive rate---none is above 50\%---but that ``damage pattern'' (evidence of damage to prey by predators) is, overall, the best of the bunch, having a true-positive rate of 47.3\% and a zero false-discovery rate, meaning that if this mode is observed it is a reliable indication of predation.  The latter initially appears to be an interesting and informative statement (it also applies to the ``co-occurrence'' and ``body size'' modes), but in fact it is not as useful as it might at first seem.  The zero false-discovery rate occurs because this type of observation is usually seen in concert with other modes of evidence for the same edge, which together cause the algorithm to (correctly) conclude that the edge is present with high probability.  Thus we can indeed reliably infer that an edge is present when this mode is observed, but even in the absence of this mode we would probably infer the same thing in most cases.

Among the other modes, ``coprolites'' and ``track or domicile'' have the lowest true-positive rates.  And, perhaps surprisingly, ``gut contents,'' which Dunne~\etal\ consider the gold standard for establishing predation, is relatively poor on both measures, with a true-positive rate of only 16.8\% (meaning observations in this mode are rare) and a false-discovery rate of 26.9\% (meaning over a quarter of observations of this type turn out to be wrong).  This makes gut contents the second-most unreliable mode of observation, after ``morphology.''  The explanation for this result is essentially the opposite of that for ``damage pattern'' above: in a significant fraction of cases gut contents is the only type of observation in favor of an interaction.  If an interaction truly exists then the probability that none of the other modes of evidence would be observed is low.  When no other modes are observed, therefore, the algorithm concludes that there is a chance that the interaction does not in fact exist, and hence that the gut contents data are not wholly reliable.

\subsection{Individual node errors}
\label{sec:individual}
For our final example we consider the model of Section~\ref{sec:directed} in which the network is undirected but observations of it are directed and there are individual and potentially different error rates for each node.  This model is particularly appropriate for acquaintance network data.

Suppose we have a social network of friendship or acquaintance and the structure of the network is measured by surveying people and asking them who their friends are.  We use the configuration model as our network model, with the parameters~$\omega$ and~$\phi_i$ being given once again by Eqs.~\eqref{eq:cmomega} and~\eqref{eq:cmphi}.  For our data model we use a variant of the approach described in Section~\ref{sec:directed} and define $\alpha_{ik}$ to be the probability that individual~$i$ identifies another individual as a friend if there are $k$ (undirected) edges between them in the ground-truth network.  Then the data likelihood given the ground truth is
\begin{align}
P(D|\mat{A},\alpha) &= \prod_{i\ne j} \prod_{k=0}^\infty
  \bigl[ \alpha_{ik}^{E_{ij}} (1-\alpha_{ik})^{N_{ij}-E_{ij}}
  \bigr]^{\delta_{k,A_{ij}}} \nonumber\\
  &\hspace{1em}{}\times \prod_i \prod_{k=0}^\infty
  \bigl[ \alpha_{ik}^{E_{ii}/2} (1-\alpha_{ik})^{(N_{ii}-E_{ii})/2}
  \bigr]^{\delta_{k,A_{ii}}},
\end{align}
where $E_{ij}$ is the number of times (out of $N_{ij}$ total) that $i$ identifies $j$ as a friend (which under normal circumstances will be either zero or one) or twice that number when $i=j$.

\begin{widetext}
Following the same lines of argument as previously, we then find that
\begin{equation}
\alpha_{ik} = {\sum_j Q_{ij}(k) E_{ij}\over\sum_j Q_{ij}(k) N_{ij}},
\qquad
Q_{ij}(k) = {(\omega\phi_i\phi_j)^k/k!
             \bigl[\alpha_{ik}^{E_{ij}} (1-\alpha_{ik})^{N_{ij}-E_{ij}}
          \alpha_{jk}^{E_{ji}} (1-\alpha_{jk})^{N_{ji}-E_{ji}}\bigr] \over
          \sum_{k=0}^\infty (\omega\phi_i\phi_j)^k/k!
             \bigl[\alpha_{ik}^{E_{ij}} (1-\alpha_{ik})^{N_{ij}-E_{ij}}
          \alpha_{jk}^{E_{ji}} (1-\alpha_{jk})^{N_{ji}-E_{ji}}\bigr]}.
\end{equation}
\end{widetext}
As with the model of Section~\ref{sec:cmindep}, it will in the common case of a sparse network usually be adequate to compute only $Q_{ij}(1)$ and assume $Q_{ij}(0) = 1 - Q_{ij}(1)$ and $\hat{A}_{ij} = Q_{ij}(1)$, all other probabilities $Q_{ij}(k)$ with $k\ge2$ being negligible.

\begin{figure*}
\begin{center}
\includegraphics[width=13cm]{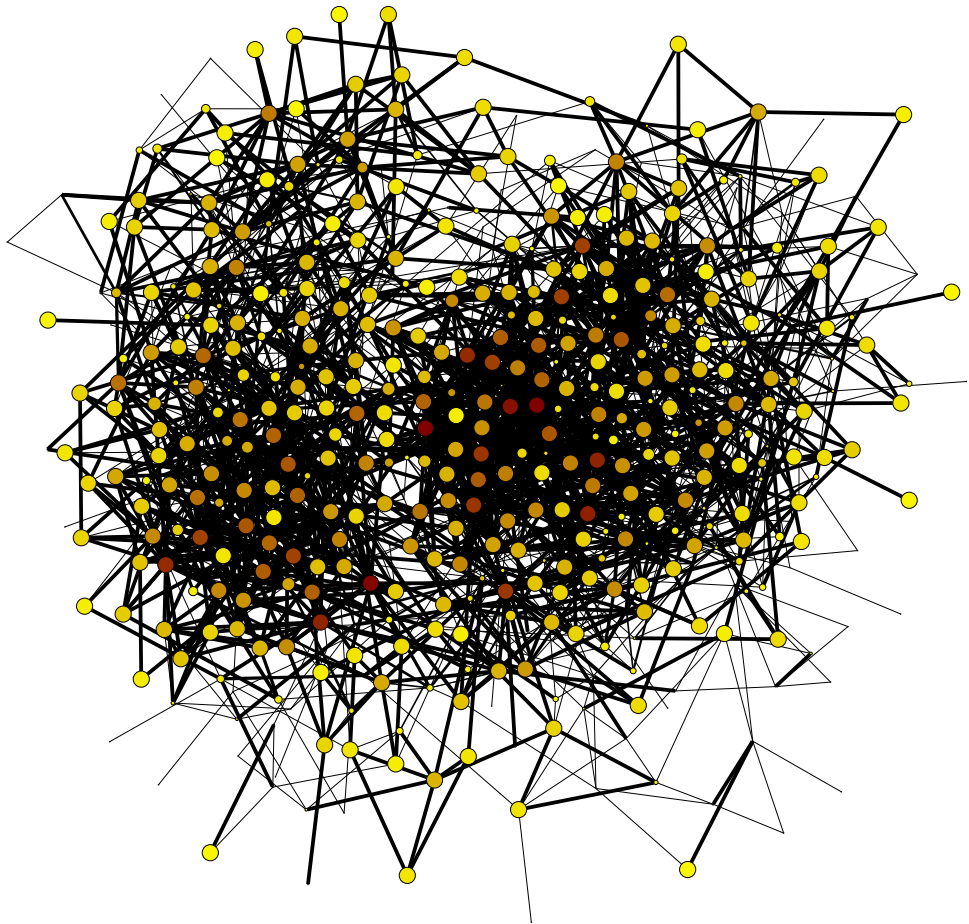}
\end{center}
\caption{The inferred network of friendships among the students in a medium-sized American high school and its feeder middle school.  Edge thicknesses represent the inferred posterior probabilities of the edges.  Node sizes represent the inferred values of the degree parameters~$\phi_i$ for the configuration model.  Node colors represent the estimated average precision of reports made by the corresponding individual, precision being the probability that a reported friendship actually exists.  Colors range from yellow (lowest precision) to red (highest precision).  Only edges that are reported to exist by at least one participant are shown, and nodes with no reported edges are omitted.}
\label{fig:school}
\end{figure*}

\subsection{Example application}
As an example of this algorithm we consider data from the US National Longitudinal Study of Adolescent Health~\cite{Resnick97,AddHealth}, known colloquially as the ``Add Health'' study, a large-scale study of students in US middle and high schools conducted during the 1990s.  Among other things the study asked students to identify their friends, but it was found that individuals often disagreed about friendships: as discussed in Section~\ref{sec:directed}, a substantial fraction of all claims of friendship are unreciprocated, implying a significant level of false positives, false negatives, or both in the data, and we can estimate these levels by applying our methods.

There were 84 populations surveyed in the Add Health study, where a population consisted of a high school and an associated feeder middle school.  The populations ranged in size from dozens of students to thousands and the methods described here could be applied to any of them.  In Fig.~\ref{fig:school} we show results for a medium-sized population with 542 students.  In this figure the widths of the edges once again vary to indicate the estimated probabilities~$Q_{ij}$.  In addition we vary the diameters of the nodes in proportion to the estimated degree parameter~$\phi_i$, which can be thought of as a measure of the sociability or popularity of individuals.  We also vary the colors of the nodes to denote the reliability of their reports of friendships.  As our measure of reliability we use the \defn{precision}, which is the probability that a reported friendship is actually correct.  As with the false-discovery rate of Section~\ref{sec:foodweb}, an expression for the precision can be written using Bayes' rule:
\begin{align}
P(A_{ij}=1|E_{ij}=1) &= P(E_{ij}=1|A_{ij}=1) {P(A_{ij}=1)\over P(E_{ij}=1)}
  \nonumber\\
  &= {\omega\phi_i\phi_j\alpha_i\over\omega\phi_i\phi_j\alpha_i+\beta_i}.
\end{align}
The numbers we use to compute the colors of the nodes are the average value of this precision over all the friendships an individual reports.

The figure reveals a network with a dense core of strongly connected nodes (perhaps divided into two parts), plus a sparser periphery of more weakly connected nodes.  Most nodes appear to have roughly the same value of~$\phi_i$ (they appear about the same size), though a small subset seem to be ``less sociable'' (they appear smaller).  Most nodes also have relatively low precision (the lighter end of the color scale); only a handful, mostly in the interior of the figure, fall in the high-precision range (darker colors).

\section{Conclusions}
In this paper we have developed a class of expectation-maximization (EM) algorithms that allow one to infer the structure of an observed network from noisy, error-prone, or incomplete measurements.  These algorithms take raw observational data concerning the structure of networks and return a posterior probability distribution over possible structures the network could take.  This posterior distribution can then be used to estimate any other network quantity of interest along with the standard error on that estimate.  In addition our algorithms also return values for a range of model parameters that quantify the mapping between the true structure of the network and the observed data, such as true- and false-positive rates for observation of individual edges.  In many cases these parameters are of interest in their own right.

We have given three examples of practical applications of our methods to previously published network data sets.  In the first example we inferred the structure of a social network from repeated observations of physical proximity between pairs of people.  In the second example we looked at a food web data set of predator-prey interactions among a group of species.  Connections in the network are measured using a number of different techniques and, though none of techniques is very reliable, our methods allow us to combine them to make an estimate of the structure of the network.  Our third example focused again on a social network, in this case of declared friendships between students in a US high school and middle school.  In addition to allowing us to infer the structure of the friendship network, our algorithm in this case also gives us a measure of how accurately each student reports their own friendships.

One thing we have not done in this paper is look in detail at the case introduced in Section~\ref{sec:sbm} of networks that are generated from the stochastic block model (or its degree-corrected variant introduced in Section~\ref{sec:cm}).  Application of these models would allow us to simultaneously infer both the structure of the network and its division into communities.  We leave these developments, however, for future work.

\begin{acknowledgments}
The author thanks Elizabeth Bruch, George Cantwell, Jennifer Dunne, Travis Martin, Gesine Reinert, and Maria Riolo for useful comments.  This work was funded in part by the US National Science Foundation under grants DMS--1407207 and DMS--1710848.
\end{acknowledgments}

\end{document}